\documentclass[11pt]{article}
\usepackage{amssymb,amsthm,amscd,array}
\usepackage{graphics}
\usepackage[final]{epsfig}

\let\ssection=\section
\renewcommand{\section}{\setcounter{equation}{0}\ssection}

\newcommand{\bbR}{\mathbb{R}}

\newcommand{\bbC}{\mathbb{C}}

\newcommand{\cB}{{\mathcal{B}}}

\newcommand{\cK}{{\mathcal{K}}}
\newcommand{\cL}{{\mathcal{L}}}

\newcommand{\Tr}{{\mathrm{Tr}}}

\newcommand{\cF}{{\mathcal{F}}}

\newcommand{\SL}{\mathrm{SL}}
\newcommand{\SSp}{\mathrm{Sp}}
\newcommand{\Sl}{\mathrm{sl}}

\newcommand{\osp}{\mathrm{osp}}
\newcommand{\OSp}{\mathrm{OSp}}

\newcommand{\cE}{\mathcal{E}}

\newcommand{\half}{\frac{1}{2}}

\chardef\s=110
\chardef\g=103

\begin{document}

\newtheorem{theorem}{Theorem}
\newtheorem{lemma}{Lemma}[section]
\newtheorem{cor}[lemma]{Corollary}
\newtheorem{conj}[lemma]{Conjecture}
\newtheorem{proposition}[lemma]{Proposition}
\newtheorem{rmk}[lemma]{Remark}
\newtheorem{exe}[lemma]{Example}
\newtheorem{defi}[lemma]{Definition}

\def\a{\alpha}
\def\b{\beta}
\def\d{\delta}
\def\g{\gamma}
\def\om{\omega}
\def\r{\rho}
\def\s{\sigma}
\def\t{\tau}
\def\vfi{\varphi}
\def\vr{\varrho}
\def\l{\lambda}
\def\L{\Lambda}
\def\m{\mu}

\title{Supertransvectants and symplectic geometry}

\author{H. Gargoubi
\thanks{
I.P.E.I.T., 2 Rue Jawaher Lel Nehru, Monfleury \_ 1008
Tunis,
TUNISIE; hichem.gargoubi@ipeit.rnu.tn,}
\and
V. Ovsienko
\thanks{
CNRS,
Institut Camille Jordan,
Universit\'e Claude Bernard Lyon 1,
21 Avenue Claude Bernard,
69622 Villeurbanne Cedex,
FRANCE;
ovsienko@math.univ-lyon1.fr
}}

\date{}

\maketitle

\begin{abstract}
The $1|1$-supertransvectants are the $\osp(1|2)$-invariant bilinear operations
on weighted densities on the supercircle $S^{1|1}$, the projective
version of $\bbR^{2|1}$.
These operations are analogues of the famous Gordan transvectants
(or Rankin-Cohen brackets).
We prove that supertransvectants coincide with the iterated
Poisson and ghost Poisson brackets on $\bbR^{2|1}$ and
apply this result to construct star-products.
\end{abstract}

\thispagestyle{empty}

%%%%%%%%%%%%%%%%%%%%%%%%%%%%%%%%%%%%%%%%%%
%%%%%%%%%%%%%%%%%%%%%%%%%%%%%%%%%%%%%%%%%%
\section{Introduction}
%%%%%%%%%%%%%%%%%%%%%%%%%%%%%%%%%%%%%%%%%%
%%%%%%%%%%%%%%%%%%%%%%%%%%%%%%%%%%%%%%%%%%

%%%%%%%%%%%%%%%%%%%%%%%%%%%%%%%%%%%%%%%%%%
%%%%%%%%%%%%%%%%%%%%%%%%%%%%%%%%%%%%%%%%%%
\subsection{The transvectants and linear Poisson bracket: 
recapitulation}\label{TrS}
%%%%%%%%%%%%%%%%%%%%%%%%%%%%%%%%%%%%%%%%%%
%%%%%%%%%%%%%%%%%%%%%%%%%%%%%%%%%%%%%%%%%%

Consider the space, denoted by $\cF_\l$, of smooth
(complex valued) functions on $S^1$ equipped with the following
$\SL(2,\bbR)$-action:
\begin{equation}
\label{LFTwei}
f(x)\mapsto
f\left({ax+b \over cx+d}\right){(cx+d)}^{-2\l},
\end{equation}
where $x$ is the affine coordinate and
$\l\in\bbC$ is a parameter.
Note that the space $\cF_\l$ is naturally identified
with the space of weighted densities of degree $\l$
($\l$-densities for short) via
$\varphi=f(x)\,(dx)^\l$;
the action (\ref{LFTwei}) is then the standard action of
fraction-linear coordinate transformations. 

Classification of $\SL(2,\bbR)$-invariant bilinear differential
operators on $S^1$
acting in the spaces $\cal F_\lambda$
is a famous classical result of the
invariant theory. For every $k=1,2,3,\ldots$, there exists the
$\SL(2,\bbR)$-invariant bilinear differential operator
$$
J_k^{\l,\m}:
\cF_\l\otimes\cF_\m\to\cF_{\l+\m+k}
$$
given by the following explicit formula
\begin{equation}
\label{transvectant}
J_k^{\l,\m}(f,g)=
\sum_{i+j=k}
(-1)^{i}\,
{2\l+k-1\choose{}j}
{2\m+k-1\choose{}i}\,
f^{(i)}\,g^{(j)},
\end{equation}
where $f^{(i)}(x)={d^i f(x) \over dx^i}$ and where
${a\choose{}i}=\frac{a(a-1)\cdots(a-i+1)}{i!}$. The operators
(\ref{transvectant}), called {\it transvectants}, were found in 1885 by Gordan
\cite{Gor}; for almost all $(\l,\m)$, these operators are unique
$\SL(2,\bbR)$-invariant bilinear differential operators on $S^1$ acting in the
spaces $\cal F_\lambda$.  Note that one can also assume $\l$
(half)integer and consider holomorphic functions on the upper
half-plane $\cal{H}$.

Transvectants have been rediscovered by Rankin \cite{Ran} and Cohen
\cite{Coh} in the theory of modular forms and by Janson and Peetre
\cite{JP} in differential projective geometry. Zagier \cite{Zag}
(see also \cite{OS}) noticed the coincidence between the
Rankin-Cohen brackets and Gordan's transvectants. It was shown in
\cite{EG} that the transvectants are in one-to-one correspondence
with singular (i.e., vacuum or highest weight) 
vectors in the tensor product of
two Verma modules over $\Sl(2, \bbC)$.

The best way to understand the operators (\ref{transvectant}) is,
perhaps, to rewrite them in terms 
of the projective symplectic geometry, as in
\cite{Ov1} and \cite{OT}. Consider the plane $\bbR^2$ with 
coordinates $(p,q)$ and the standard symplectic form
$\omega=dp\wedge{}dq$ and the Poisson bracket
$$
\{F,G\}=
\frac{\partial F}{\partial p}\,\frac{\partial G}{\partial q}-
\frac{\partial F}{\partial q}\,\frac{\partial G}{\partial p}.
$$
The symmetry group of linear transformations in this case is the
group $\SSp(2,\bbR)\simeq\SL(2,\bbR)$. It is easy to describe all
the $\SSp(2,\bbR)$-invariant bilinear differential operators on
$C^\infty(\bbR^2)$. For every positive integer $k$, there exists a
bilinear differential operator of order $2k$ given by the
differential binomial of the form
\begin{equation}
\label{MoyTerEq}
B_k(F,G):=
\sum_{i+j=k}
(-1)^{i}\,
{k\choose{}i}\,
\frac{\partial^k F}{\partial p^i\partial q^j}\,
\frac{\partial^k G}{\partial p^j\partial q^i}.
\end{equation}
The operators $B_k$ are, in fact, iterations of the Poisson bracket
in the following sense. Consider the operator $B$ on
$C^\infty(\bbR^2)\otimes{}C^\infty(\bbR^2)$ given by
$$
B(F\otimes{}G)=
\frac{\partial F}{\partial p}\otimes{}\frac{\partial G}{\partial q}-
\frac{\partial F}{\partial q}\otimes{}\frac{\partial G}{\partial p}
$$
and the natural projection $\Tr(F\otimes{}G)=FG$. Then obviously
$$
B_k=\Tr\circ{}B^k.
$$

The expression (\ref{MoyTerEq}) is, of course, much simpler than
(\ref{transvectant}); in particular, it is independent of $\l$ and
$\m$. Nevertheless, these operators coincide up to a multiple.
Identify the space $\cF_\l(S^1)$ and the space of functions on
$\bbR^2\setminus\{0\}$ homogeneous of degree $-2\l$ by
\begin{equation}
\label{identif} 
\textstyle f(x)\longmapsto F_f(p,q)=
p^{-2\l}\,f\left( \frac{q}{p} \right),
\end{equation}
so that the affine coordinate is chosen as $x=q/p$.
\begin{exe}
 {\rm 
a) In the case $\l=-1$, the above formula identifies the 3-dimensional
space spanned by $\{1,x,x^2\}$ and the space of quadratic
polynomials spanned by $\{p^2,pq,q^2\}$; this gives two realizations
of $\mathrm{sl}(2)$: in terms of vector fields on $S^1$ and Hamiltonian
vector fields on $\bbR^2$, respectively.

b) In the case $\l=-\half$, one identifies affine functions
$1,x$ with linear functions $p,q$.
}
\end{exe}

The following observation was made in \cite{Ov1}.
\begin{proposition}
\label{FPr} One has: $ B_k(F_f, F_{g})=k!\,
F_{J_k^{\l,\m}(f,g)}. $
\end{proposition}

A simple corollary of Proposition \ref{FPr} is the fact that
the operators (\ref{transvectant}) can be used to construct an
$\SL(2,\bbR)$-invariant star-product on $T^*S^1$
(see \cite{CMZ}, \cite{Ov1}, \cite{OMMY} and \cite{OT}).
Another application of the same idea leads to a multi-dimensional
generalization of the transvectants as $\SSp(2n,\bbR)$-invariant
bilinear differential operators
on the sphere $S^{2n-1}$, see \cite{OT}.
Simple expression (\ref{MoyTerEq}) allows one to avoid any
non-trivial combinatorics.

\begin{rmk}
{\rm Formula (\ref{identif}) is somewhat mysterious, but it has a
geometric sense. Every vector field on $S^1$ admits a unique
``symplectic lift'' to a homogeneous Hamiltonian vector field on
$\bbR^2\setminus\{0\}$ and (\ref{identif}) is the unique lifting of
weighted densities commuting with the vector fields lift (cf.
\cite{OT}). }
\end{rmk}

%%%%%%%%%%%%%%%%%%%%%%%%%%%%%%%%%%%%%%%%%%
%%%%%%%%%%%%%%%%%%%%%%%%%%%%%%%%%%%%%%%%%%
\subsection{The $1|1$-supertransvectants}
%%%%%%%%%%%%%%%%%%%%%%%%%%%%%%%%%%%%%%%%%%
%%%%%%%%%%%%%%%%%%%%%%%%%%%%%%%%%%%%%%%%%%

We define the supercircle $S^{1|1}$ in terms of its superalgebra of
functions:
$C_\bbC^\infty(S^{1|1})=C_\bbC^\infty(S^{1})\,[\xi]$,
where $\xi$ is an odd (Grassmann) coordinate, i.e., $\xi^2=0$
and $x\xi=\xi{}x$.
In other words, this is the algebra of polynomials 
(of degree $\leq1$) in $\xi$ with coefficients in $C_\bbC^\infty(S^{1})$:
$$
f(x,\xi)=f_0+\xi\,f_1
$$
where $f_0,f_1$ are smooth functions on $S^1$.
The parity function $\s$ is defined 
on homogeneous in $\xi$ functions by setting
$\s(f_0(x))=0$ and $\s(\xi\,f_1(x))=1$.

%%%%%%%%%%%%%%%%%%%%%%%%%%%%%%%%%%%%%%%%%%
\subsubsection*{The fractional-linear transformations}
%%%%%%%%%%%%%%%%%%%%%%%%%%%%%%%%%%%%%%%%%%

The action of the supergroup $\OSp(1|2)$ on $S^{1|1}$ is given by
the fraction-linear transformations
$$
(x,\xi)\mapsto \left( \frac{ax+b+\g\xi}{cx+d+\d\xi},\,
\frac{\a{}x+\b+e\xi}{cx+d+\d\xi} \right),
$$
where $ ad-bc-\a\b=1, e^2+2\g\d=1, \a{}e=a\d-c\g$ and 
$\b{}e=b\d-d\g$ (cf. \cite{CMZ,DM}).

We denote by $\cF_\l$ the superspace of functions
$C_\bbC^\infty(S^{1|1})$ equipped with the following
$\OSp(1|2)$-action
\begin{equation}
\label{ActonFunct} f(x,\xi)\mapsto f\left(
\frac{ax+b+\g\xi}{cx+d+\d\xi},\, \frac{\a{}x+\b+e\xi}{cx+d+\d\xi}
\right) \left( cx+d+\d\xi \right)^{-2\l},
\end{equation}
where $\l\in\bbC$ is a parameter.

As usual, it is much easier to deal with the
infinitesimal version of this action. The action of the
orthosymplectic Lie superalgebra $\osp(1|2)$ on $S^{1|1}$
corresponding to the $\OSp(1|2)$-action is spanned by three even and
two odd vector fields:
$$
\begin{array}{l}
\displaystyle \osp(1|2)_0=\mathrm{Span}\left(
\frac{\partial}{\partial{}x}, \qquad
x\frac{\partial}{\partial{}x}+\half\,\xi\,\frac{\partial}{\partial\xi},
\qquad
x^2\frac{\partial}{\partial{}x}+x\xi\,\frac{\partial}{\partial\xi}
\right),
\\[14pt]
\displaystyle \osp(1|2)_1=\mathrm{Span}\left( D, \qquad x\,D
\right),
\end{array}
$$
where
$$
D=\frac{\partial}{\partial\xi}+
\xi\frac{\partial}{\partial{}x}
$$
is an odd vector field satisfying
$\half\,[D,D]=\frac{\partial}{\partial{}x}$.

The action of $\osp(1|2)$ on $\cF_\l$ corresponding to the group
action (\ref{ActonFunct}) is easy to calculate:
\begin{equation}
\label{LieDac}
\begin{array}{l}
\displaystyle L^\l_{\frac{\partial}{\partial{}x}}
=\frac{\partial}{\partial{}x}, \qquad
L^\l_{x\frac{\partial}{\partial{}x}}
=x\frac{\partial}{\partial{}x}+\l, \qquad
L^\l_{x^2\frac{\partial}{\partial{}x}+
x\xi\,\frac{\partial}{\partial\xi}}
=x^2\frac{\partial}{\partial{}x}+
x\xi\,\frac{\partial}{\partial\xi}, +2\l\,x
\\[14pt]
\displaystyle L^\l_D=D, \qquad L^\l_{x\,D}= x\,D+2\l\xi
\end{array}
\end{equation}
which is nothing but the Lie derivative of $\l$-densities (see,
e.g., \cite{GMO}).

\begin{rmk}
{\rm Note that the odd elements $D$ and $xD$ generate the whole
$\osp(1|2)$ so that an operator commuting with the action of these
two elements commutes with the $\OSp(1|2)$-action. }
\end{rmk}

We will also use the following odd vector field on $S^{1|1}$
$$
\overline{D}= \frac{\partial}{\partial\xi}-
\xi\,\frac{\partial}{\partial{}x},
$$
which defines the {\it contact structure} on $S^{1|1}$
since it spanns the kernel of the contact 1-form $\a=dx+\xi\,d\xi$, 
see \cite{Lei,GMO,DM} (Manin \cite{Man} calls this vector field the
canonical SUSY-structure)
\footnote{For an invariant description of the operators $D$ and
$\overline{D}$, in physical papers denoted by $Q$ and $D$,
respectively, see \cite{Shch}.}.
It is characterized by the relations for the Lie superbrackets
$$
[D,\overline{D}]=0, 
\qquad 
\half\,[\overline{D},\overline{D}]=
-\frac{\partial}{\partial{}x}.
$$
An important property of $\overline{D}$ is that this vector field
is invariant (up to multiplication by functions)
under the $\OSp(1|2)$-action.
In particular, one
has $[xD,\overline{D}]=-\xi\overline{D}$.

Every differential operator on $S^{1|1}$ can be expressed in terms of
$\overline{D}$.
For instance, one has for the partial derivatives:
$$
\frac{\partial}{\partial{}x}=
-\overline{D}^2,
\qquad
\frac{\partial}{\partial\xi}=
\overline{D}-\xi\,\overline{D}^2.
$$

%%%%%%%%%%%%%%%%%%%%%%%%%%%%%%%%%%%%%%%%%%
\subsection{Supertransvectants: an explicit formula}
%%%%%%%%%%%%%%%%%%%%%%%%%%%%%%%%%%%%%%%%%%

The {\it supertransvectants} are the bilinear $\OSp(1|2)$-invariant
maps $ J^{\l,\m}_{k}: \cF_\l\otimes\cF_\m \to\cF_{\l+\m+k} $ where
$k=0,\frac{1}{2},1,\frac{3}{2},2,\ldots$. The supertransvectants
were introduced by Gieres and Theisen in \cite{GT} and \cite{Gie},
see also \cite{Hua}. Their (slightly modified) explicit formula is
\begin{equation}
\label{SupTransGeFor} J^{\l,\m}_{k}(f,g)= \sum_{i+j=2k} C^k_{i,j}\,
\overline{D}^{i}(f)\,\overline{D}^{j}(g),
\end{equation}
where the numeric coefficients are
\begin{equation}
\label{SupTransMainFor} C^k_{i,j}= (-1)^{\left(
\left[\frac{j+1}{2}\right]+j(i+\s(f)) \right)}\, \frac{ \left(
\begin{array}{c}
\left[k\right]\\[4pt]
\left[\frac{2j+1+(-1)^{2k}}{4}\right]
\end{array}
\right) \left(
\begin{array}{c}
2\l+\left[k-\frac{1}{2}\right]\\[4pt]
\left[\frac{2j+1-(-1)^{2k}}{4}\right]
\end{array}
\right) } { \left(
\begin{array}{c}
2\m+\left[\frac{j-1}{2}\right]\\[4pt]
\left[\frac{j+1}{2}\right]
\end{array}
\right) },
\end{equation}
where $[a]$ denotes the integer part of $a\in\bbR$. 
It can be checked directly that these operators are,
indeed, $\OSp(1|2)$-invariant.

%%%%%%%%%%%%%%%%%%%%%%%%%%%%%%%%%%%%%%%
\subsection{Comments}
%%%%%%%%%%%%%%%%%%%%%%%%%%%%%%%%%%%%%%%

It is an interesting feature of the supersymmetric case,
that the operators
labeled by integer $k$ are \textit{even}, and by semi-integer $k$
are \textit{odd}.

The two first examples of the supertransvectants, namely for
$k=\half$ and $k=1$, play a particular role. These operations are
not only $\OSp(1|2)$-invariant, but also invariant with respect to
the full infinite-dimensional conformal Lie superalgebra $\cK(1)$
(also known as the centerless Neveu-Schwarz algebra); for a complete
description of bilinear invariant $\cK(N)$-operators for $n=1$, 2
and 3 over contact vector fields with {\it polynomial} coefficients,
see \cite{LKV} and \cite{Lei}. The first-order supertransvectant
$J_1$ is nothing but the well-known contact bracket on $S^{1|1}$. 
The odd supertransvectant $J_\half$ also belongs to the list of
invariant operators from
\cite{LKV} and \cite{Lei}, but this operator is much less known.
We will show that this operator defines a very interesting
operation of ``antibracket'' on the $\cK(1)$-modules of densities.

%%%%%%%%%%%%%%%%%%%%%%%%%%%%%%%%%%%%
\subsection{The main results}
%%%%%%%%%%%%%%%%%%%%%%%%%%%%%%%%%%%%

The main purpose of this paper is to give an interpretation
of the supertransvectants in terms of the linear symplectic
superspace $\bbR^{2|1}$ with coordinates $(p,q,\t)$ and the standard
symplectic form $\om=dp\wedge{}dq+d\t\wedge{}d\t$. This
interpretation considerably simplifies the explicit expression of
the supertransvectants and their definition. It also allows one to
apply some algebraic constructions of Poisson geometry, as
star-products and suggests multi-dimensional generalizations of the
supertransvectants.

The standard Poisson bracket on $\bbR^{2|1}$ is given by
\begin{equation}
\label{SpO}
\{F,G\}=
\frac{\partial F}{\partial p}\,\frac{\partial G}{\partial q}-
\frac{\partial F}{\partial q}\,\frac{\partial G}{\partial p}+
\frac{\partial F}{\partial \t}\,\frac{\partial G}{\partial \t}.
\end{equation}
Consider the space of functions on $\bbR^{2|1}$
with singularities at $(p,q)=(0,0)$
satisfying the condition
$
\cE(F)=2\,F,
$
where
$$
\cE=p\,\frac{\partial}{\partial{}p}+
q\,\frac{\partial}{\partial{}q}+
\t\,\frac{\partial}{\partial\t}
$$
is the Euler field;
such functions are called homogeneous of degree $2$.
This space is stable with respect to the bracket (\ref{SpO}),
therefore, it is
a Lie (but not Poisson) superalgebra.
This is nothing but the conformal superalgebra $\cK(1)$.

We introduce one more, odd, operation on
$C^\infty(\bbR^{2|1})$:
\begin{equation}
\label{Ghost}
\{F,G\}_{\rm gPb}=\frac{\partial F}{\partial \t}\,\cE(G)-
(-1)^{\s(F)}\,\cE(F)\,\frac{\partial G}{\partial \t}
+\t\,
\left(
\frac{\partial F}{\partial p}\,\frac{\partial G}{\partial q}-
\frac{\partial F}{\partial q}\,\frac{\partial G}{\partial p}
\right),
\end{equation}
where $\s$ is the parity function.
We call it the \textit{ghost Poisson bracket}.

We will study the geometric and algebraic meaning of operation
(\ref{Ghost}).
Its crucial property is $\cK(1)$-invariance.

\begin{theorem}
\label{XXX}
The ghost bracket (\ref{Ghost}) is invariant with respect
to the action of the conformal algebra $\cK(1)$.
\end{theorem}

It turns out that the Poisson bracket
restricted to the homogeneous functions coincides with the
supertransvectant $J_1$,
while the ghost Poisson bracket coincides with $J_\half$.
In the framework of deformation quantization,
we will consider ``iterated'' Poisson brackets
(\ref{SpO}) and (\ref{Ghost}).

\begin{theorem}
\label{YYY}
The supertransvectants $J_k$ with integer $k$ coincide with
the iterated Poisson bracket (\ref{SpO}), while those with
semi-integer $k$ are obtained by the iteration of
(\ref{SpO}) with (\ref{Ghost}).
\end{theorem}

To the best of our knowledge, operations of type (\ref{Ghost}) have
not been studied (see \cite{Yve} for a survey of algebraic
structures in Poisson geometry and \cite{LKV} for that in
supergeometry). Note that (\ref{Ghost}) is not invariant with
respect to the full Poisson superalgebra
$\left(C^\infty(\bbR^{2|1}),\{\,,\,\}\right)$.

%%%%%%%%%%%%%%%%%%%%%%%%%%%%%%%%%%%%
\subsection{Open problems}
%%%%%%%%%%%%%%%%%%%%%%%%%%%%%%%%%%%%

Grozman, Leites and Shchepochkina
listed all simple Lie superalgebras of vector fields on the
supercircles \cite{GLS1}
(it is instructive to compare their list with that in
\cite{Kac}), and thus indicated the scope of work for
possible superizations of Gordan's transvectants. The case we
consider is the first on the agenda.
Although there are four infinite series and several exceptional cases
of simple stringy (or superconformal) superalgebras, there are only
7 (or, perhaps, 12: this has to be investigated) among them that contain
the subalgebra of fraction linear transformations similar to the
projective actions of $\mathrm{sl}(2)=\mathrm{sp}(2)$ or $\osp(1|2)$
considered here.

%%%%%%%%%%%%%%%%%%%%%%%%%%%%%%%%%%%%%%%%%%
%%%%%%%%%%%%%%%%%%%%%%%%%%%%%%%%%%%%%%%%%%
\section{The Poisson bracket and the ghost bracket}
%%%%%%%%%%%%%%%%%%%%%%%%%%%%%%%%%%%%%%%%%%
%%%%%%%%%%%%%%%%%%%%%%%%%%%%%%%%%%%%%%%%%%

Let us consider the first examples of supertransvectants:
$J^{\l,\m}_\half$ and $J^{\l,\m}_1$ .
To simplify the notations, throughout this section, we denote
these operators by $(\,,\,)$ and $[\,,\,]$, respectively.

%%%%%%%%%%%%%%%%%%%%%%%%%%%%%%%%%%%%%%%%%%
\subsection{The two operations}
%%%%%%%%%%%%%%%%%%%%%%%%%%%%%%%%%%%%%%%%%%

The supertransvectant of order $\half$ is
\begin{equation}
\label{HalfOrd}
(f,g)=
\m\,\overline{D}(f)\,g-
(-1)^{\s(f)}\,\l\,f\,\overline{D}(g).
\end{equation}
This odd operator is extremely interesting.
We will show that it is invariant with respect to the full
infinite-dimensional superconformal algebra
(and therefore has a geometric meaning).

The first-order supertransvectant is
\begin{equation}
\label{FirstOrd}
[f,g]=
\m\,f'\,g-
\l\,f\,g'-
(-1)^{\s(f)}\,\frac{1}{2}\,
\overline{D}(f)\,\overline{D}(g).
\end{equation}
This even operation is nothing but the well-known Poisson
bracket on $S^{1|1}$
(see, e.g., \cite{Lei}, \cite{LKV} and also \cite{GMO}).

%%%%%%%%%%%%%%%%%%%%%%%%%%%%%%%%%%%%%%%%%%
\subsection{The Poisson superalgebra $\cF$
and the conformal superalgebra $\cK(1)$}
%%%%%%%%%%%%%%%%%%%%%%%%%%%%%%%%%%%%%%%%%%

Consider the continuous sum (direct integral) of all spaces $\cF_\l$:
$$
\cF=\cup_{\l\in\bbC}\cF_\l,
$$
the collection of operations $J^{\l,\m}_1$ defines a bilinear map
$[\,,\,]:\cF\otimes\cF\to\cF$.

\begin{lemma}
\label{JacLem}
The operation $J_1$ defines the structure of a Poisson Lie superalgebra on
$\cF$.
\end{lemma}
\begin{proof}
Straightforward.
\end{proof}

%%%%%%%%%%%%%%%%%%%%%%%%%%%%%%%%%%%%%%%%%%
%\subsection{The conformal superalgebra $\cK(1)$ and its modules}
%%%%%%%%%%%%%%%%%%%%%%%%%%%%%%%%%%%%%%%%%%

The space $\cF_{-1}\subset\cF$ is a Lie subalgebra
since it is stable with respect to the bracket (\ref{FirstOrd}).
This is precisely the conformal superalgebra on $S^{1|1}$,
also known as the Lie superalgebra of contact vector fields
(see \cite{Lei},\cite{GLS1} and also \cite{GMO}),
or the (centerless) Neveu-Schwarz algebra.
Let us denote this Lie subalgebra $\cK(1)$.
Each space $\cF_\l$ is a $\cK(1)$-module.

%%%%%%%%%%%%%%%%%%%%%%%%%%%%%%%%%%%%%%%%%%
\subsection{Invariance of the supertransvectant $J_\half$}
%%%%%%%%%%%%%%%%%%%%%%%%%%%%%%%%%%%%%%%%%%

The operation (\ref{HalfOrd}) is an additional,
odd, bracket on the superspace $\cF$.
The crucial property of this ghost bracket is that it
is \textit{invariant} with respect to the action of
the conformal subalgebra $\cK(1)\subset\cF$.

\begin{proposition}
\label{HalfInvPro}
The operation (\ref{HalfOrd}) on $\cF$ is $\cK(1)$-invariant.
\end{proposition}
\begin{proof}
One has to check that for $f\in\cF_{-1}$ and arbitrary
$g\in\cF_\m$ and $h\in\cF_\nu$ one has
\begin{equation}
\label{InvC}
[f,(g,h)]
=
(-1)^{\s(f)}\,
([f,g],h)+
(-1)^{\s(f)(\s(g)+1)}\,
(g,[f,h]).
\end{equation}
It can be done by straightforward calculation.
Note however, that the identity (\ref{InvC}) is a particular case of
Theorem \ref{XXX} whose proof will be given in Section \ref{PSec}.
\end{proof}

%%%%%%%%%%%%%%%%%%%%%%%%%%%%%%%%%%%%%%%%%%
\subsection{The algebraic structure on $\cF_{-\half}$}
%%%%%%%%%%%%%%%%%%%%%%%%%%%%%%%%%%%%%%%%%%

The $\cK(1)$-module $\cF_{-\half}$ is a ``square root'' of
$\cK(1)\cong\cF_{-1}$.
This space is stable with respect
to the operation $(\,,\,)$.
Adopting the basis
$$
V_n=x^{n+\half},
\qquad
\Psi_n=\xi\,x^{n},
$$
one obtains explicitly
\begin{equation}
\label{GhosRel}
\begin{array}{rcl}
(V_n,V_m) &=&
\left(m-n\right)\Psi_{n+m},\\[8pt]
(\Psi_n,V_m) &=&
V_{n+m}=-(V_m,\Psi_n),\\[8pt]
(\Psi_n,\Psi_m) &=&
2\Psi_{n+m}.
\end{array}
\end{equation}

\begin{proposition}
The algebra
$(\cF_{-\half},(\,,\,))$
satisfies the following four properties:

\begin{enumerate}
\item
the odd part $\left(\cF_{-\half}\right)_1$ is
a commutative associative subalgebra;

\item
the odd part $\left(\cF_{-\half}\right)_1$ acts on the even
part $\left(\cF_{-\half}\right)_0$ by
$\rho_{\psi}v:=(\psi,v)$ and one has
$$
\rho_{\vfi}\circ\rho_{\psi}
+\rho_{\psi}\circ\rho_{\vfi}=
\rho_{(\vfi,\psi)}
$$
for all $\vfi,\psi\in\left(\cF_{-\half}\right)_1$;

\item
the map
$(\,,\,):\left(\cF_{-\half}\right)_0
\otimes\left(\cF_{-\half}\right)_0\to
\left(\cF_{-\half}\right)_1$
is anti-symmetric
and $\left(\cF_{-\half}\right)_1$-invariant, namely
$$
\rho_{\psi}(v,w)=
(\rho_{\psi}v,w)+(v,\rho_{\psi}w)
$$
for all $\psi\in\left(\cF_{-\half}\right)_1$ and
$v,w\in\left(\cF_{-\half}\right)_0$;

\item
the Jacobi identity is satisfied:
$$
(u,(v,w))+(v,(w,u))+(w,(u,v))=0,
$$
for all $u,v,w\in\left(\cF_{-\half}\right)_0$.
\end{enumerate}

\end{proposition}
\begin{proof}
The properties 1--4 of can be checked
directly.
\end{proof}

%%%%%%%%%%%%%%%%%%%%%%%%%%%%%%%%%%%%%%%%%%
%%%%%%%%%%%%%%%%%%%%%%%%%%%%%%%%%%%%%%%%%%
\section{The symplectic lifting}
%%%%%%%%%%%%%%%%%%%%%%%%%%%%%%%%%%%%%%%%%%
%%%%%%%%%%%%%%%%%%%%%%%%%%%%%%%%%%%%%%%%%%

In this section we show that the supertransvectants (\ref{FirstOrd})
and (\ref{HalfOrd}) coincide with the Poisson bracket (\ref{SpO})
and the ghost bracket (\ref{Ghost}). We prove Theorem \ref{XXX}.

%%%%%%%%%%%%%%%%%%%%%%%%%%%%%%%%%%%%%%%%%%
\subsection{Homogeneous functions}
%%%%%%%%%%%%%%%%%%%%%%%%%%%%%%%%%%%%%%%%%%

Let us define a symplectic lifting of the space $\cF$.
To any function $f\in\cF_\l$ we associate a function on
$\bbR^{2|1}$ homogeneous of degree $-2\l$.
The explicit formula is
$f(x,\xi)\mapsto{}F_f(p,q,\t)$, where
\begin{equation}
\label{SymLift}
\textstyle
F_f(p,q,\t)=
p^{-2\l}\,f\left(
\frac{q}{p},\,\frac{\t}{p}
\right)\equiv
p^{-2\l}\,f_0\left(
\frac{q}{p}\right)+
\t\,p^{-2\l-1}\,f_1\left(
\frac{q}{p}\right)
\end{equation}
and where $(p,q,\t)$ are coordinates on $\bbR^{2|1}$.
Abusing the notations, from now on, we will also denote
$\cF_\l$ the space of homogeneous functions on $\bbR^{2|1}$
of degree $-2\l$.

This lifting is invariant in the following sense.

\begin{proposition}
(i)
The 1-transvectant $J^{\l,\m}_1$, see the explicit
formula (\ref{FirstOrd}), corresponds
to the Poisson bracket (\ref{SpO}):
$$
F_{[f,g]}=
\half\,\{F_f,F_g\},
$$

(ii)
The $\half$-transvectant (\ref{HalfOrd}) corresponds
to the odd bracket (\ref{Ghost}):
$$
F_{(f,g)}=
-\half\,\{F_f,F_g\}_{\rm gPb}.
$$
\end{proposition}
\begin{proof}
Just substitute the expression
(\ref{SymLift}) to (\ref{SpO}) and (\ref{Ghost})
and compare the result with (\ref{FirstOrd}) and (\ref{HalfOrd}).
\end{proof}

A nice feature of the lifting (\ref{SymLift}) is
that it intertwines
the standard embedding of $\osp(1|2)$ into the Poisson
algebra given by the quadratic polynomials
$$
\osp(1|2)=
\mathrm{Span}
\left(
p^2,\,pq,\,q^2;\quad\t{}p,\,\t{}q
\right)
$$
with the $\osp(1|2)$-action (\ref{LieDac}).
Again, the odd elements $\t{}p,$ and $\t{}q$ generate the
whole algebra.

\begin{rmk}
{\rm
The lifting (\ref{SymLift}) has a similar geometric
meaning as that of (\ref{identif}).
The Lie superalgebra
$\cK(1)\cong\cF_{-1}$ corresponds
to the space of functions on $\bbR^{2|1}$ homogeneous
of degree 2 and formula (\ref{SymLift}) is the unique
way to identify weighted densities with homogeneous
functions that intertwines (\ref{FirstOrd}) and (\ref{SpO}).
}
\end{rmk}

%%%%%%%%%%%%%%%%%%%%%%%%%%%%%%%%%%%%%%%%%%
\subsection{Invariance of the ghost Poisson bracket}\label{PSec}
%%%%%%%%%%%%%%%%%%%%%%%%%%%%%%%%%%%%%%%%%%

Let us prove Theorem \ref{XXX}.

To show that the ghost bracket (\ref{Ghost}) is invariant
with respect to the action of $\cK(1)$,
one has to show that
$$
\{F,\{G,H\}_{\rm gPb}\}
=
(-1)^{\s(F)}\,
\{\{F,G\},H\}_{\rm gPb}+
(-1)^{\s(F)(\s(G)+1)}\,
\{G,\{F,H\}\}_{\rm gPb}
$$
for every function $F\in\cF_{-1}$.
To do this, we adopt the technique routine in Poisson geometry.
The bracket (\ref{Ghost}) is given by the following
``ghost Poisson'' bivector
\begin{equation}
\label{LaMb}
\Lambda=
\frac{\partial}{\partial\t}\wedge\cE+
\t\,P,
\end{equation}
where
$
P=\frac{\partial}{\partial{}p}
\wedge\frac{\partial}{\partial{}q}
$
is the even part of the Poisson bivector.
The equivariance condition is equivalent to the
fact that the Hamiltonian
vector field, $X_F$, with respect to the Poisson bracket
(\ref{SpO}) preserves the bivector $\Lambda$
that can be readily checked.
\qed

\begin{rmk}
{\rm
There is a uniqueness statement.
It follows from the classification of the supertransvectants,
that, for generic $(\l,\m)$,
the ghost bracket (\ref{Ghost}) is a unique odd
bilinear homogeneous map $\cF_\l\otimes\cF_\m\to\cF_\nu$
commuting with the $\cK(1)$-action.
}
\end{rmk}

%%%%%%%%%%%%%%%%%%%%%%%%%%%%%%%%%%%%%%%%%%
%%%%%%%%%%%%%%%%%%%%%%%%%%%%%%%%%%%%%%%%%%
\section{Supertransvectants from the symplectic viewpoint}
%%%%%%%%%%%%%%%%%%%%%%%%%%%%%%%%%%%%%%%%%%
%%%%%%%%%%%%%%%%%%%%%%%%%%%%%%%%%%%%%%%%%%

In this section we prove Theorem \ref{YYY}.
We realize the supertransvectants
in terms of the iterated brackets
(\ref{SpO}) and (\ref{Ghost}).
As a corollary of this result,
we construct a star-product involving
the supertransvectants $J_k$ as $k$-th order terms.

%%%%%%%%%%%%%%%%%%%%%%%%%%%%%%%%%%%%%%%%%%
\subsection{Even supertransvectants as the iterated Poisson bracket}
%%%%%%%%%%%%%%%%%%%%%%%%%%%%%%%%%%%%%%%%%%

Consider the linear operator $\cB$ acting on the space
$C^\infty(\bbR^{2|1})\otimes{}C^\infty(\bbR^{2|1})$ given by
\begin{equation}
\label{BMap}
\cB(F\otimes{}G)=
\frac{\partial F}{\partial p}\otimes{}\frac{\partial G}{\partial q}
-\frac{\partial F}{\partial q}\otimes{}\frac{\partial G}{\partial p}
+\frac{\partial F}{\partial \t}\otimes{}\frac{\partial G}{\partial \t}
\end{equation}
The Poisson bracket (\ref{SpO}) is given by the composition:
$\{\,,\,\}=\Tr\circ\cB$ where $\Tr$ is the operator of projection
$\Tr(F\otimes{}G)=FG$.

Define the ``iterated Poisson brackets''
$\cB_k=\Tr\circ\cB^k$, with $k=1,2,\ldots$.
One readily gets the explicit formula:
\begin{equation}
\label{iterPoisson}
\cB_k(F,G)=
B_k(F,G)+k\,B_{k-1}
\left(\frac{\partial F}{\partial \t},\frac{\partial G}{\partial \t}
\right),
\end{equation}
where $B_k$ is the iterated bracket (\ref{MoyTerEq}) on $\bbR^2$.

\begin{proposition}
\label{PrIt}
The iterated Poisson bracket (\ref{iterPoisson})
is $\osp(1|2)$-invariant for every integer $k$.
\end{proposition}
\begin{proof}
The $\osp(1|2)$-action on $\bbR^{2|1}$ is generated by
two odd elements: $\t{}p$ and $\t{}q$.
Let us check that
$$
\{t{}p,\cB_k(F,G)\}=
\cB_k(\{t{}p,F\},G)+
(-1)^{\s(F)}\,\cB_k(F,\{t{}p,G\}).
$$
If $F,G$ are even then the above relation is evident.
For $F$ even and $G=\t{}G_1$ odd one has the condition
$$
p\,B_k(F,G_1)
=
k\,B_{k-1}
\left(
\frac{\partial F}{\partial q},G_1
\right)+B_k(F,p\,G_1)
$$
that follows from formula (\ref{MoyTerEq}).
Finally, for $F=\t{}F_1,G=\t{}G_1$, one gets the relation:
$$
k\t\,\left(B_{k-1}(F_1,G_1)\right)_q
=
\t\,\left(B_k(pF_1,G_1)-B_k(F_1,pG_1)\right)
$$
which is obviously true.
\end{proof}

The bilinear map $\cB_k$ restricted to the homogeneous
functions defines the map
$$
\cB_k:\cF_\l\otimes\cF_\m\to
\cF_{\l+\m+k}
$$
which is $\osp(1|2)$-invariant.
It follows then from the uniqueness of the supertransvectants
that the maps
$J_k^{\l,\m}$ and $\cB_k|_{\cF_\l\otimes\cF_\m}$ are proportional.
Taking particular functions $p^{-2\l}$ and $q^{-2\m}$,
one now checks that the proportionality coefficient
is $2^k\,k!$ and finally
\begin{equation}
\label{PST}
F_{J_k^{\l,\m}(f,g)}=
\frac{1}{2^k\,k!}\,\cB_k(F_f,F_g).
\end{equation}
for generic, and therefore, for all $(\l,\m)$.

%%%%%%%%%%%%%%%%%%%%%%%%%%%%%%%%%%%%%%%%%%
\subsection{Iterated ghost Poisson bracket and the odd supertransvectants}
%%%%%%%%%%%%%%%%%%%%%%%%%%%%%%%%%%%%%%%%%%

Define an analogous linear operator corresponding to the
ghost bracket (\ref{Ghost}) by the following formula:
\begin{equation}
\label{GMap}
\begin{array}{rcl}
\displaystyle
\Game(F\otimes{}G) &=&
\frac{\partial F}{\partial \t}\otimes\cE(G)-
(-1)^{\s(F)}\,\cE(F)\otimes{}\frac{\partial G}{\partial \t}\\[12pt]
&&
+
\displaystyle
\chi(f,g)\,
\left(
\t{}\frac{\partial F}{\partial p}\otimes{}\frac{\partial G}{\partial q}
-\t{}\frac{\partial F}{\partial q}\otimes{}\frac{\partial G}{\partial p}
+\frac{\partial F}{\partial p}\otimes\t{}\frac{\partial G}{\partial q}
-\frac{\partial F}{\partial q}\otimes\t{}\frac{\partial G}{\partial p}
\right)
\end{array}
\end{equation}
where $\chi(f,g)$ is a function depending on the parity
of $f$ and $g$:
$$
\chi(f,g)=
\half+\frac{\left(
1+(-1)^{(\s(f)+1)(\s(g)+1)}
\right)}{4}
$$
Clearly
$$
\{\,,\,\}_{\rm gPb}=\Tr\circ\Game.
$$

Let us define the odd iterated brackets:
\begin{equation}
\label{OdIt}
\cB_{k+\half}=\Tr\circ\Game\circ\cB^k
\end{equation}
for $k=1,2,\ldots$.

\begin{proposition}
\label{OdDBItP}
The odd brackets $\cB_{k+\half}$ are $\osp(1|2)$-invariant.
\end{proposition}
\begin{proof}
Similar to the proof of Proposition \ref{PrIt}.
\end{proof}
Again, the proportionality coefficient can be
calculated:
\begin{equation}
\label{PSTGhost}
F_{J_{k+\half}^{\l,\m}(f,g)}=
-\frac{1}{2^k\,k!}\,\cB_{k+\half}(F_f,F_g).
\end{equation}

\begin{rmk}
{\rm
(i)
The definition (\ref{OdIt}) does not depend on
the order of composition of the operators
$\Game$ and $\cB$ since one has
$$
\Tr\circ\Game\circ\cB^k=
\Tr\circ\cB^\ell\circ\Game\circ\cB^m,
$$
for $\ell+m=k$.

(ii)
the map (\ref{GMap}) is the ``square root''
of the map (\ref{BMap}) in the following sense:
$$
\Tr\circ\Game^2
=\half
\left(
(1+(-1)^{\s(F)(\s(G)+1)})\,
(\m+1)-
(1+(-1)^{\s(G)(\s(F)+1)})\,
(\l+1)
\right)
\Tr\circ\cB,
$$
when restricted to the homogeneous functions
$\cF_\l\otimes{}\cF_\m$.
}
\end{rmk}

%%%%%%%%%%%%%%%%%%%%%%%%%%%%%%%%%%%%%%%%%%
\subsection{An $\osp(1|2)$-invariant star-product}
%%%%%%%%%%%%%%%%%%%%%%%%%%%%%%%%%%%%%%%%%%

The coincidence (\ref{PST}) defines
a pull-back of the standard Moyal-Weyl star-product
on $\bbR^{2|1}$ to
an invariant star-product on the Poisson algebra $\cF$.
The explicit formula is very simple:
\begin{equation}
\label{StAr}
f*g=f\,g+
\sum_{k=1}^\infty\,t^k\,
J_k^{\l,\m}(f,g),
\end{equation}
for all $f\in\cF_\l$ and $g\in\cF_\m$. The operation (\ref{StAr}) is
an associative product on the space of formal series $\cF[[t]]$
which is a deformation of the standard commutative product of
functions. The star-product (\ref{StAr}) is obviously
$\osp(1|2)$-invariant.

Note that the operation (\ref{StAr}) involves only even
supertransvectants.
It would be interesting to understand if there is another
deformation that contains the odd terms as well.

%%%%%%%%%%%%%%%%%%%%%%%%%%%%%%%%%%%%%%%%%%
%%%%%%%%%%%%%%%%%%%%%%%%%%%%%%%%%%%%%%%%%%
\section{Appendix}
%%%%%%%%%%%%%%%%%%%%%%%%%%%%%%%%%%%%%%%%%%
%%%%%%%%%%%%%%%%%%%%%%%%%%%%%%%%%%%%%%%%%%

For the sake of completeness, let us give here a proof of
the fact that, for the generic $(\l,\m)$, the supertransvectants
(\ref{SupTransGeFor}) with coefficients (\ref{SupTransMainFor})
are the unique $\osp(1|2)$-invariant bidifferential operators.

An arbitrary bidifferential operator can be written in the form
(\ref{SupTransGeFor}) with coefficients
$C_{i,j}\in{}C^\infty(S^{1|1})$. The action of a vector field $X$
on the operator (\ref{SupTransGeFor}) is then given by
$$
\cL(B)(f,g)
:=
\sum_{i+j=k}
C_{i,j}\left(
L_{X}(\overline{D}^i)(f)\,\overline{D}^j(g)
+(-1)^{i+\s(f)} \overline{D}^i(f)\,L_{X}(\overline{D}^j)(g)
\right)
$$
We will use the generators $D$ and $xD$ of $\osp(1|2)$. The
invariance condition with respect to the first generator $D$ proves
that each $C_{i,j}$ is an even constant. Consider the vector field
$xD$. First, we calculate the action of $xD$ on the operators
$\overline{D}^i:\cF_\l\to\cF_\m$. One has
$$
\begin{array}{rcl}
L_{xD}\,(\overline{D}^{2p+1})
&:=&
(xD+2\m\xi)\,\overline{D}^{2p+1}
+
\overline{D}^{2p+1}\,(xD+2\l\xi)\\[10pt]
&=&
(2\l+p)\,\overline{D}^{2p}
+(2\m-2\l-2p-1)\,\xi\overline{D}^{2p+1}
\end{array}
$$
for $i=2p+1$ and
$$
\begin{array}{rcl}
L_{xD}\,(\overline{D}^{2p})
&:=&
(xD+2\m\xi)\,\overline{D}^{2p}
-
\overline{D}^{2p}\,(xD+2\l\xi)\\[10pt]
&=&
p\,\overline{D}^{2p-1}+(2\m-2\l-2p)\,\xi\overline{D}^{2p}
\end{array}
$$
for $i=2p$.
In particular, if $\m=\l+{i\over2}$, one obtains
$$
L^{\l,\l +{i\over 2}}_{xD}\,(\overline{D}^{i})
=
\left\{
\begin{array}{ll}
(2\l + {(i-1)\over 2})\,\overline{D}^{i-1}
& \hbox{if $i$ is odd},\\[8pt]
{i\over 2}\,\overline{D}^{i-1}
& \hbox{if $i$ is even.}
\end{array}
\right.
$$

The equivariance equation, $L_{xD}(J)(f,g)=0$,
for a bidifferential operator
$J$ gives now the following system:
\begin{equation}
\label{system1}
\begin{array}{rcl}
(2\l+l)\,C_{2l+1,2m}
&=&
- (-1)^{\s(f)} (2\m+m)\,C_{2l,2m+1}\\[10pt]
l\,C_{2l,2m-1}
&=&
(-1)^{\s(f)} m\,C_{2l-1,2m} \\[10pt]
(2\l+l)\,C_{2l+1,2m-1}
&=&
- (-1)^{\s(f)} m\,C_{2l,2m} \\[10pt]
l\,C_{2l,2m}
&=&
(-1)^{\s(f)} (2\m+m)\,C_{2l-1,2m+1}
\end{array}
\end{equation}

Explicit solution of the system (\ref{system1})
leads to the following critical (or ``resonant'')
set
$$
\textstyle
I_k=
\left\{
0,-{1\over2},-1,-{3\over2}\dots,
-{1\over2}
\left[
{k-1\over2}
\right]
\right\}
$$
and one has to separate the following four cases.

1)
If $\l,\m\not\in I_k$,
then the system (\ref{system1}) has a unique
(up to a multiplicative constant) solution given by
(\ref{SupTransMainFor}).

2)
If one of the weights $\l$ or $\m$
belongs to $I_k$
but the second one does not, then the system (\ref{system1})
has a unique (up to a multiplicative constant) solution.
If, say, $\l=\frac{1-m}{4}$ for some odd $m$, then
the corresponding bilinear $\osp(1|2)$-invariant operator
is given by
$$
f\otimes{}g\longmapsto
J^{\frac{1+m}{4},\m}_{k-m}
\left(\overline{D}^{m}(f),g
\right).
$$

3) If $\l=\frac{1-m}{4}$ for some odd $m$ and
$\m=\frac{1-\ell}{4}$ for some odd $\ell$,
and if $\ell+m>k$,
then the solution is still unique and is of the form
$$
f\otimes{}g\longmapsto
J^{\frac{1+m}{4},\m}_{k-m}
\left(\overline{D}^{m}(f),g
\right)=
J^{\l,\frac{1+\ell}{4}}_{k-\ell}
\left(f,\overline{D}^{\ell}(g)
\right).
$$

4)
If $\l=\frac{1-m}{4}$ for some odd $m$ and
$\m=\frac{1-\ell}{4}$ for some odd $\ell$,
and if $\ell+m>k$,
then there are two independent solutions
$$
f\otimes{}g\longmapsto
J^{\frac{1+m}{4},\m}_{k-m}
\left(\overline{D}^{m}(f),g
\right)=
J^{\l,\frac{1+\ell}{4}}_{k-\ell}
\left(f,\overline{D}^{\ell}(g)
\right).
$$
and
$$
f\otimes{}g\longmapsto
J^{\frac{1+m}{4},\frac{1+\ell}{4}}_{k-m-\ell}
\left(\overline{D}^{m}(f),\overline{D}^{\ell}(g)
\right).
$$

\vskip 1cm

\textbf{Acknowledgements}. It is a great pleasure to thank Christian
Duval  and Dimitry Leites for numerous enlightening discussions and help. 
We are also
grateful to Charles Conley, Fran\c{c}ois Gieres,
Claude Roger and Serge Parmentier for their valuable comments at
different stages of this work.

%%%%%%%%%%%%%%%%%%%%%%%%%%%%%%%%%%%%%%%%%%%%%%%%%%%%%%%%%%%%%%%%%%%%%%%%%%%%%%
%%%%%%%%%%%%%%%%%%%%%%%%%%%%%%%%%%%%%%%%%%%%%%%%%%%%%%%%%%%%%%%%%%%%%%%%%%%%%%

\end{document}